# Magnitude and phase reciprocity calibration of ultrasonic piezoelectric disk in air

Kenneth K. Andersen[1,3], Per Lunde[1,2], Jan Kocbach[2]

[1] University of Bergen, Dept. of Physics and Technology, PB 7803, NO-5020 BERGEN, Norway
[2] Christian Michelsen Research AS, PB 6031 Postterminalen, NO-5892 BERGEN, Norway
[3] University College of Southeast Norway, PB 235, 3603 Kongsberg, Norway
Contact email: kenneth.andersen@hbv.no

## Abstract

A modified conventional three-transducer reciprocity calibration method is used to measure the magnitude and phase responses of the transmitting voltage response and the free-field open-circuit receiving voltage sensitivity of ultrasonic piezoelectric transducers radiating in air at 1 atm. The transducers used in this work are 20 x 2 mm circular piezoelectric ceramic disks with their first and second radial modes at approximately 100 and 250 kHz, respectively. The transducer characterization is supported and aided by finite element simulations of the measurement system and the measured frequency responses. Preliminary results indicate that the magnitude and phase responses of the transmitting voltage response and the free-field open-circuit receiving voltage sensitivity can be measured with fair accuracy in a limited frequency band around the first radial mode of the piezoelectric ceramic disk. Further work is needed to demonstrate and quantify the accuracy actually obtained using the three-transducer reciprocity calibration method, to achieve transducer characterization at a calibration accuracy level, and to achieve calibration above the calibrated frequency range of the B&K 4138 microphone.

## 1 Introduction

Ultrasonic measurement technologies are used in a wide range of application areas and industries, from petroleum and marine applications, to medicine. For gas applications, such as fiscal measurement [1–3], quality, and energy measurement of natural gas [4–6], accurate calibration data for ultrasonic transducers may be important to frequencies of 300 kHz and higher.

A range of calibration methods are available for transducers in gas or liquid [7–25]. In gas, challenges concerning these calibration techniques arise for frequencies exceeding approximately 150 kHz. For instance, for the 1/8 inch B&K 4138 condenser microphone, accurate calibration data are available from the manufacturer up to 140 kHz [26, 27].

The present work addresses characterization of transducers for use in gas at ultrasonic frequencies, as a contribution in the long-time perspective to achieve accurate magnitude and phase characterization at these frequencies, possibly at calibration accuracy levels.

A three-transducer reciprocity calibration method is considered for measurement of the transmitting voltage response and the free-field open-circuit receiving voltage sensitivity of a transducer radiating in air at 1 atm. Piezoelectric ceramic disks are used as the transmitting and receiving transducers, to simplify finite element (FE) analysis of the measurement setup and the measured magnitude and phase responses. Simulations are used

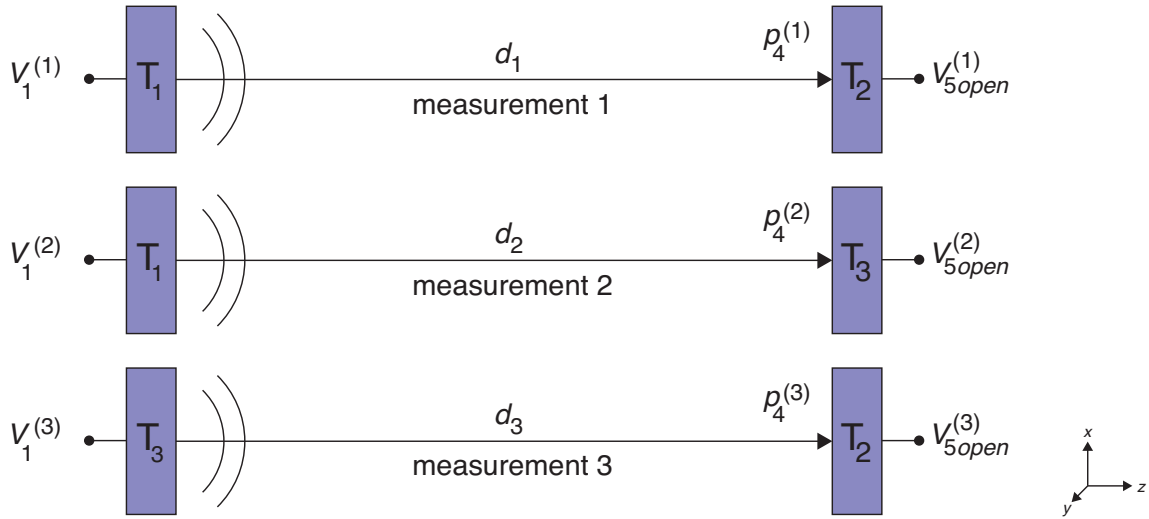

Figure 1: Schematic of the three measurements needed to perform calibration by the reciprocity calibration method.

(i) to aid in control and improvement of the measurements, (ii) for comparison with and interpretation of the measurement results, (iii) to calculate necessary correction factors to the measurements, and (iv) to determine a 360° ambiguity in the measured phase response at low frequencies, at which poor signal-to-noise ratio (SNR) is experienced.

Preliminary measurement and simulation results are shown for the magnitude and phase responses of (i) the open-circuit loss-free transmit-receive voltage-to-voltage transfer function, (ii) the transmitting voltage response, and (iii) the free-field open-circuit receiving voltage sensitivity, for circular piezoelectric ceramic disks vibrating at frequencies up to 300 kHz, a frequency band covering the first and second radial mode of the disk. Promising characterization results are obtained in the frequency range 75–125 kHz around the first radial mode at about 100 kHz, in which a SNR beyond 40 dB is achieved.

The present paper represents an update from [28], based on the work in [29], and building on prior work such as refs. [30–37].

## 2 Measurement theory

### 2.1 Reciprocity calibration method

To perform calibration by the reciprocity method three transducers and three measurements are needed, cf. Fig. 1 where $T_1$ acts as a transmitter, $T_2$ acts as a receiver, $T_3$ acts as a reciprocal transducer [1] [2], $p_4$ is the on-axis free- and far-field sound pressure, $V_1$ is the input voltage at the terminals of the transmitter, $V_{5open}$ is the open-circuit output voltage at the terminals of the receiver, and $d_1$, $d_2$ and $d_3$ are the separation distances between the transmitter and receiver for measurement 1, 2, and 3, respectively. A superscript is imposed on the quantities $V_1$, $p_4$ and $V_{5open}$ to denote what measurement the quantity is obtained from, and the subscripts refer to where the quantities exist in the measurement model, cf. Fig. 2.

---

[1] $T_3$ is a passive, linear and reversible electromechanical or electroacoustical transducer such that coupling is equal in either direction [38]

[2] It is assumed that the receivers are in the far-field of the transmitting transducer.

From the measurement set-up in Fig. 1, $T_1$ can be calibrated as a transmitter of sound, and $T_2$ can be calibrated as a receiver of sound, i.e. the calibrated quantities are the complex transmitting voltage response [7–9, 34] [3]

$$S_V^{T_1} = \left[ \frac{1}{J^{(3)} Z_{T_3}} \frac{H_{15open}^{VV(1)} H_{15open}^{VV(2)}}{H_{15open}^{VV(3)}} \frac{d_1}{d_0} \frac{d_2}{d_3} e^{-ik^{(1)}(d_0-d_1)} e^{-ik^{(2)}(d_0-d_2)} e^{ik^{(3)}(d_0-d_3)} \right]^{\frac{1}{2}}, \quad (1)$$

and the complex receiving voltage sensitivity [7–9, 34]

$$M_V^{T_2} = \left[ J^{(3)} Z_{T_3} \frac{H_{15open}^{VV(1)} H_{15open}^{VV(3)}}{H_{15open}^{VV(2)}} \frac{d_1}{d_0} \frac{d_3}{d_2} e^{-ik^{(1)}(d_0-d_1)} e^{ik^{(2)}(d_0-d_2)} e^{-ik^{(3)}(d_0-d_3)} \right]^{\frac{1}{2}}, \quad (2)$$

where $J^{(3)}$ is the spherical reciprocity parameter for measurement 3, $Z_{T_3}$ is the electrical input impedance of the reciprocal transducer $T_3$, $k^{(n)} = 2\pi f / c$ is the wave number for the $n$'th measurement, $f$ is frequency, $c$ is the speed of sound, and

$$H_{5open}^{VV(n)} \equiv \frac{V_{15open}^{(n)}}{V_1^{(n)}}, \quad (3)$$

is the complex open-circuit loss-free voltage-to-voltage transfer function relating the input voltage at the transmitter to the open-circuit output voltage at the receiver for the $n$'th measurement, and $n = 1, 2$ or 3 denotes the measurement number.

## 2.2 Open-circuit loss-free voltage-to-voltage transfer function

To perform the three measurements in Fig. 1 additional measurement equipment is needed. In Fig. 2 a schematic of the complete measurement set-up is given. Each block represents the physical equipment, or the environment, used to conduct the measurements, and in Appendix B.1, Table 1 a description of the node voltages are given.

The signal propagation is in two branches. In branch (1) the function generator [39] is connected to the transmitter through the coaxial cable 1 (RG-178), the transmitter is connected to the receiver through the medium (air), the receiver is connected to the signal amplifier [40] and filter [41] through the coaxial cable 3 (RG-178), the signal amplifier and filter are connected to the oscilloscope [42] through the coaxial cable 4 (RG-58). In branch (2) the function generator is connected to the oscilloscope through the coaxial cable 2 (RG-58).

The oscilloscope is used to record the two signals corresponding to the two branches. The recorded signals, hereby referred to as voltages, are denoted $V_{0m} = |V_{0m}|e^{i\theta_{0m}}$ and $V_6 = |V_6|e^{i\theta_6}$, where $|\ |$ denotes magnitude, and $\theta_{0m}$ and $\theta_6$ are the phases of $V_{0m}$ and $V_6$, respectively.

The two recorded voltages are not included in Eqs. (1) and (2), however $H_{15open}^{VV}$ is. To obtain $H_{15open}^{VV}$ corrections are applied to the recorded voltages. The corrections account for the influence the measurement equipment and connecting cables have on the signal propagating through the measurement set-up, and are given as transfer functions. To compensate for propagation losses in air and possible deviations from far-field conditions and near-field effects, two additional corrections are introduced, $C_\alpha$ and $C_{dif}$, respectively.

---

[3]Confer Appendix A for derivation of Eqs. (1) and (2).

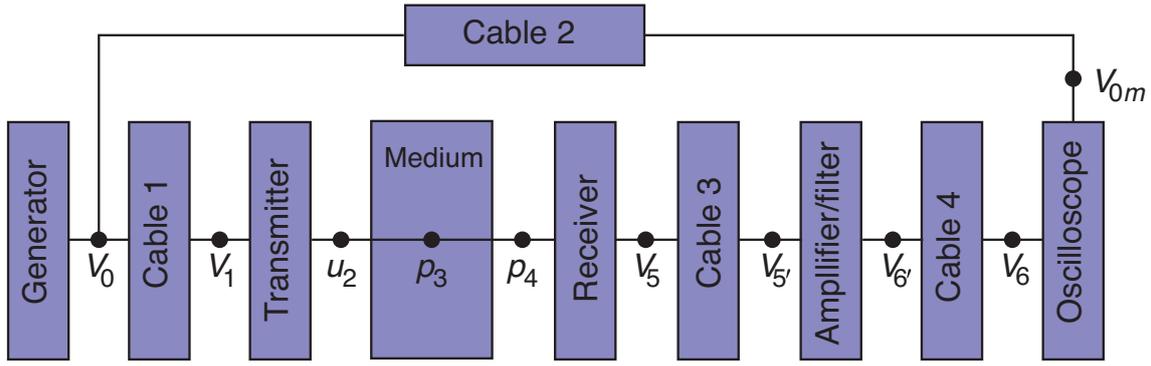

Figure 2: Model of the measurement set-up with node numbering.

A detailed description of the corrections are beyond the scope of this paper, and it is referred to e.g. [29, 31, 34, 36]. For completeness, the definitions of the corrections are given in Appendix B.2 Table 2.

When the corrections are applied to the recorded voltages the complex open-circuit loss-free transfer function is given as

$$H^{VV}_{15open} = \frac{V_6}{V_{0m}} \cdot \frac{1}{H^{VV}_{0m1} \cdot H^{VV}_{5open5'} \cdot H^{VV}_{5'6open} \cdot H^{VV}_{6open6}} \cdot C_\alpha \cdot C_{dif}, \qquad (4)$$

which can be represented as

$$H^{VV}_{15open} = |H^{VV}_{15open}|e^{i\theta_{15open}}, \qquad (5)$$

where $|H^{VV}_{15open}|$ is the magnitude and $\theta_{15open}$ is the phase of $H^{VV}_{15open}$, respectively.

The magnitude and phase of the corrections will not be discussed further, however both the phase and magnitude of the recorded voltages are the subjects of the next section, and the total expression for the phase $\theta_{15open}$ is deferred to Sec. 2.4.

## 2.3  Phase and magnitude of the recorded voltages $V_{0m}$ and $V_6$

The function generator in Fig. 2 outputs a single frequency sinusoidal burst which length is assumed long enough for steady-state conditions to be realized at the receiver. Both $V_{0m}$ and $V_6$ are recorded in the time-domain and a transformation to the frequency-domain is performed using a Fast Fourier Transform (FFT) [43]. Any transformation as such is associated with a window function, and a rectangular window, denoted FFT-window, is used in the current work. Thus, a lower and upper bound must be defined for both magnitude and phase for both voltages.

To obtain the magnitudes $|V_{0m}|$ and $|V_6|$ a steady-state region is located in the received bursts and the FFT-window's lower and upper bounds are placed in a zero crossing. The spectra are computed and only the magnitudes of the spectra are kept.

The phase of $V_{0m}$ is obtained by placing the FFT-window's lower bound in $t = 0$, where $t$ is time, and the upper bound in the steady-state region, towards the end of the burst. The spectrum is computed and only the phase $\theta_{0m}$ is kept.

To obtain the phase of $V_6$ it is useful to decompose the total phase, $\theta_6$, in two components: 1) the phase associated with the measurement equipment and transducers, denoted slowly-varying phase, $\theta_6^{slow}$, and 2) the phase associated with the acoustical wave propagation in the medium, $2\pi f t_p$. Formally, this can be stated as [43]

$$\theta_6 = \theta_6^{slow} - 2\pi f t_p, \qquad (6)$$

where $f$ is frequency, and $t_p$ is the time-of-flight of the acoustical wave propagation in air. An estimate of $t_p$ is obtained by

$$t_p = \frac{d}{c(f)}, \qquad (7)$$

where $d$ is the separation distance between the transmitter and receiver, and $c(f)$ is the speed of sound in air corrected for dispersion [44, 45], cf. Appendix D.

If the total phase, $\theta_6$, is of interest, then only an estimate of $t_p$ is needed, i.e. $t_p$ has to be within one signal period of the actual signal onset, cf. Appendix C. However, if the slowly-varying phase, $\theta_6^{slow}$, is of interest, then this becomes directly dependent on $t_p$ estimating the actual signal onset.

Estimations of $t_p$ are associated with relatively large uncertainties due to the uncertainties in determining $d$, cf. Sec. 3.1, and calculating $c(f)$ [4]. However, as an input parameter to the Eqs. (1–2) the total phase $\theta_{15open}$ is used. Thus, the total phase $\theta_6$ is used when calculating Eqs. (1–2). The consequence of this is that the uncertainty of the phases $\angle S_V^{T_1}$ and $\angle M_V^{T_2}$ are not affected by the uncertainty in either $d$ or $t_p$.

## 2.4 Phase of the open-circuit loss-free voltage-to-voltage transfer function

The phase $\theta_{15open}$ in Eq. (5) can be expressed without the exponential notation, and it can further be decomposed in a similar manner as the phase $\theta_6$, i.e.

$$\theta_{15open} = \theta_{15open}^{slow} - 2\pi f t_p, \qquad (8)$$

where $\theta_{15open}$ is the total phase of $H_{15open}^{VV}$, $\theta_{15open}^{slow}$ is the slowly-varying phase associated with the transmitter and receiver [5], and $2\pi f t_p$ is the phase associated with the acoustical wave propagation in air.

Using Eq. (4), the phases $\theta_{15open}$ and $\theta_{15open}^{slow}$ can be expressed as [29]

$$\begin{aligned}\theta_{15open} &= \theta_6 + \theta_{dif} - (\theta_{0m} + \theta_{0m1} + \theta_{5open5'} + \theta_{5'6open} + \theta_{6open6}), \\ \theta_{15open}^{slow} &= \theta_6^{slow} + \theta_{dif} - (\theta_{0m} + \theta_{0m1} + \theta_{5open5'} + \theta_{5'6open} + \theta_{6open6}),\end{aligned} \qquad (9)$$

where the phases have adopted the subscript notation from Eq. (4), e.g. the phase of $H_{0m1}^{VV}$ is denoted $\theta_{0m1}$ [6].

---

[4] Kramer [44] gives the uncertainty of the zero frequency speed of sound model to be 300 ppm. In [46] the uncertainty of the speed of sound model is indicated to be 500 ppm based on [44, 47].

[5] The phase $\theta_6^{slow}$ is associated with all of the measurement equipment, including the transmitter and receiver, however $\theta_{15open}^{slow}$ is only associated with the transmitter and receiver as corrections for the phases associated with the measurement equipment are performed.

[6] $C_\alpha$ is a real quantity and does not contribute to Eq. (9).

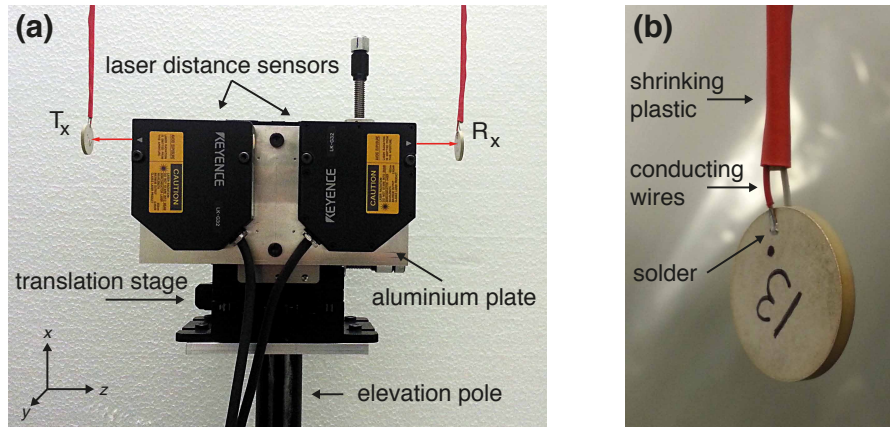

Figure 3: (a) Picture of the two laser distance sensors mounted on a manual xyz-translation stage. The laser distance sensors are shown in position for a measurement on the distance between the transmitter and receiver. (b) Picture a piezoelectric disk suspended in air.

## 3 Measurement methods

### 3.1 Measurement distance

In [29] it is shown how the separation distance between the transmitter and receiver, $d$, is obtained by measurements utilizing two laser distance sensors from Keyence [48], hereby referred to as sensors, and a linear translation stage LS270 from PI miCos [49]. In Fig. 3 (a) a picture of the two sensors is given [7]. The sensors are mounted in opposite directions on a 5 mm thick aluminum plate that is mounted to a manual xyz-translation stage from Thorlabs [50]. The two sensors and the manual xyz-translation stage are moved in position for a distance measurement using an elevation pole from Gitzo [51].

When the sensors and transmitter and receiver are placed as shown in Fig. 3 (a) the separation distance between the transmitter and receiver is approximately 0.24 m. The separation distance investigated in the current work is $d = 0.50$ m. To realize a separation distance equal to 0.50 m the receiver is moved in opposite direction of the transmitter utilizing the linear translation stage. The uncertainty associated with the measurement distance, $d$, is calculates to be ±40 μm [29].

### 3.2 Piezoelectric disks

In the current work piezoelectric disks from Meggit Sensing Systems [8] [52], of approximate diameter and thickness 20 x 2 mm, are used as both transmitters and receivers. In Fig. 3 (b) an picture of a piezoelectric disk suspended in air is shown. Visible in the figure are the two conducting wires (red and white) that are soldered onto the electrodes of the piezoelectric disk. The wires are fastened to a welding rod by shrinking plastic.

The consequence of this suspension method is that the alignment of the front face of the disks with the xy-plane is rather arbitrary. However, since the sensors are mounted on the xyz-translation stage, the sensors can be moved across the surface of the disks and deviations in the xy-plane can be measured. Thus, the disks can be re-aligned until the deviations in the xy-plane are less than a given value. In the current work the disks were re-aligned until the deviations were less than 20 μm. Re-alignment of the disks were performed by 1) loosening the screw that holds the welding rod in place and then rotating the disk, and 2) pushing on the bottom of the disk.

---
[7]The abbreviations $T_X$ and $R_X$ are used for transmitter and receiver, respectively.
[8]Previously known as Ferroperm.

Measurements have indicated that failure to co-axially align the transmitter and receiver results in fluctuating phase values (not shown here).

### 3.3 Non-linearity in piezoelectric disks

The amplitude of the excitation voltage is 10 V, however, around the series resonance frequency (~100 kHz) non-linearities are observed [29, 31, 34]. The non-linearities are attributed to the transmitting piezoelectric disk. To compensate for this, a 1 V excitation voltage is used in the frequency range 90–105 kHz. Non-linearities are expected, and observed, at the second radial mode too (~250 kHz) [31, 34], however in the current work no compensation for this has been performed. For both the first and second radial modes the non-linearities have been observed and documented using electrical input impedance measurements [29, 31, 34].

## 4 Finite element modeling

All finite element simulations are performed using FEMP 5.1, Finite Element Modeling of Ultrasonic Piezoelectric Transducers [53].

For an axisymmetric simulation problem, the simulation is defined in the $rz$-plane, where $r = \sqrt{x^2 + y^2}$, and $(x, y, z)$ are the Cartesian coordinates. The 3D solution is obtained by assuming symmetry about the $z$-axis. The simulation problem is solved using a direct harmonic analysis, with the disk immersed in a fluid. The simulation problem is divided into a region of finite elements and a region of infinite elements. The finite element region consists of the piezoelectric disk as well as the fluid loading, and is solved using 8 node isoparametric elements. The infinite element region consist of the infinite elements which are solved using 12th order conjugated Astley-Leis infinite elements. In the finite element region, seven or nine elements per shear wavelength are used in the simulations. The medium, air, is simulated without losses.

Since the material constants obtained from Ferroperm are associated with high uncertainties (as much as ± 10 percent [52]) an adjusted material data set, developed at the university of Bergen (UiB) [54], is used in the current work. However, since the material constants are not specifically adjusted for the disks used in the current work, deviations between the measurement and simulations are still expected.

## 5 Preliminary results and discussion

### 5.1 Open-circuit loss-free voltage-to-voltage transfer function, $H^{VV}_{15open}$

Three measurements corresponding to the measurements indicated in Fig. 1 are performed. The three measurements consist of different transmitter and receiver pairs, thus the measurement set-up has been dismantled in-between each measurement. The measurements are compared to a simulation of the same quantity.

In Fig. 4 (a) and (b) $|H^{VV}_{15open}|$ and $\theta^{slow}_{15open}$ are shown, respectively. The crosses indicate the interval where the 1 V measurements are used. The grey area, 70–125 kHz, indicates where a SNR > 40 dB is achieved. This frequency range corresponds to where the calibrations are expected to be performed with adequate SNR [29].

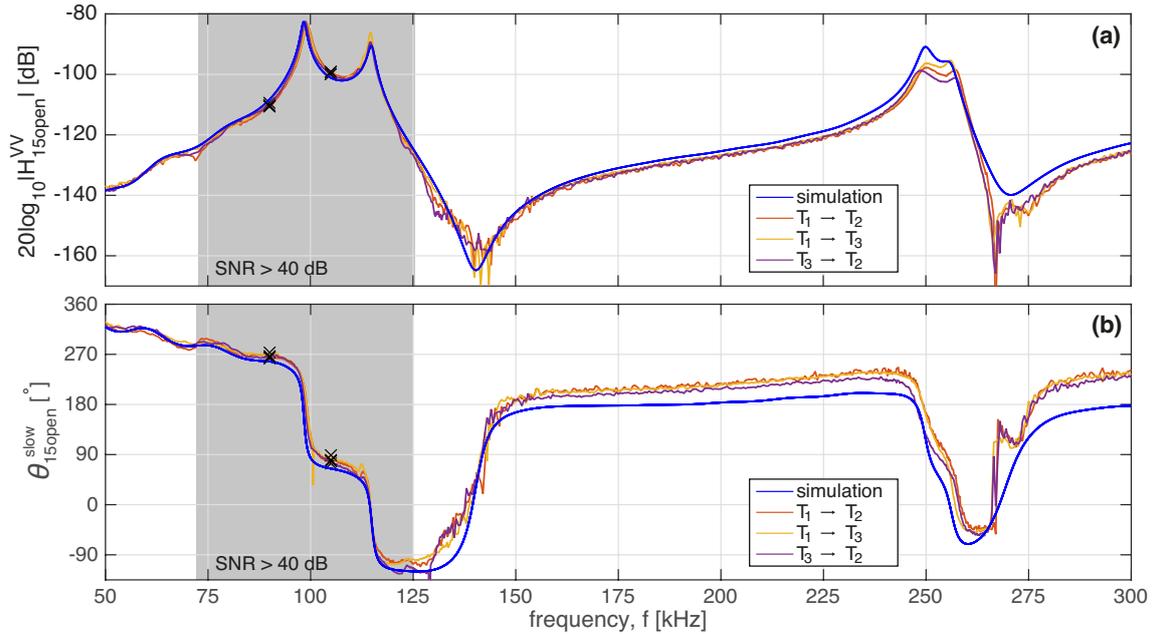

Figure 4: (a) Magnitude and (b) slowly-varying phase of the open-circuit loss-free voltage-to-voltage transfer function $H^{VV}_{15open}$. The grey area indicates the frequency range where a SNR > 40 dB is achieved.

In Fig. 4 (a) around 70–125 kHz a fair correspondence between the three measurements are observed; the observed deviations are attributed to differences in the physical properties of the piezoelectric disks. A fair correspondence is also seen between the three measurements and the simulation, and the observed differences are attributed to the uncertainties in the material parameters used in the FE-simulations [29]. The deviations around 250 kHz are partly explained by non-linearities that was not compensated for by using a lower excitation voltage, and by differences in the physical properties of the piezoelectric disks. From ~160–300 kHz a frequency dependent deviation between the three measurements and the simulation is observed. It is hypothesized, but not investigated, that this deviation might partly be due to the uncertainty in the correction factor accounting for attenuation.

In Fig. 4 (b) around 70–125 kHz a fair correspondence between the three measurements are observed. A fair correspondence between the measurements and the simulations is also achieved, where the deviations are less than 20°. Above 125 kHz a frequency dependent deviation between all three measurements and the simulation is observed; at 200 kHz a deviation of ~30° is observed, and at 300 kHz a deviation of ~60° is observed. This frequency dependent deviation is hypothesized to stem from 1) deviations in the measurement distance, $d$, compared to the actual separation distance between the transmitter and receiver, and 2) uncertainties in the estimate $t_p$ compared to the actual time-of-flight of the acoustical wave, and 3) deviations in the co-axial alignment of the disks with the $xy$-plane.

For both Fig. 4 (a) and (b) the large deviation around 260–275 kHz is attributed to diffraction effects [29].

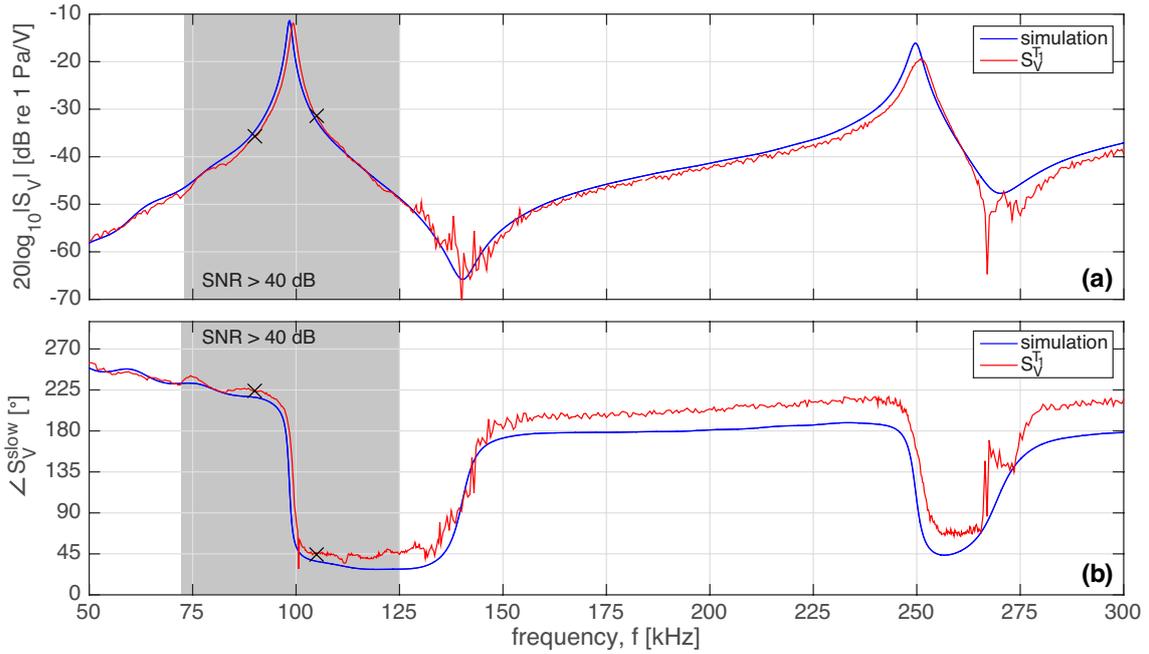

Figure 5: (a) Magnitude $|S_V|$ and (b) slowly-varying phase $\angle S_V^{slow}$ of the transmitting voltage response, $S_V^{T_1}$. The grey area indicates the frequency range where a SNR > 40 dB is achieved.

## 5.2 Transmitting voltage response, $S_V^{T_1}$

In Fig. 5 the transmitting voltage response $S_V^{T_1}$, cf. Eq. (1), is shown for (a) magnitude $|S_V|$ and (b) slowly-varying phase $\angle S_V^{slow} = \angle S_V^{T_1} + 2\pi f t_p$. The measurement is compared to a simulation of the same quantity.

In Fig. 5 (a) around 70–125 kHz a fair correspondence between the measurement and the simulation is observed. Noteworthy is the deviation around 100 kHz. The simulation peaks approximately 1 kHz before the measurement. This deviation is attributed to the material constants used in the FE-simulation [9]. The deviation around 250 kHz is partly explained by non-linearity and by the material constants used in the FE-simulations. At 200 kHz a deviation between measurement and simulations of approximately 1 dB is observed, and at 300 kHz a deviation of approximately 2 dB is observed.

In Fig. 5 (b) around 70–125 kHz a fair correspondence between the measurement and the simulation is observed, where the observed deviations are less than 10°. As in Fig. 4 (b) the deviation between the measurement and simulation increase with increasing frequency. At 200 kHz a deviation of ~20° is observed, and at 300 kHz a deviation of ~35° is observed. The deviation is hypothesized to stem from the same sources listed in Sec. 5.1.

For both Fig. 5 (a) and (b) the large deviation around 260–275 kHz is attributed to diffraction effects [29].

## 5.3 Receiving voltage sensitivity, $M_V^{T_2}$

In Fig. 6 the receiving voltage response $M_V^{T_2}$, cf. Eq. (2), is shown for (a) magnitude $|M_V|$ and (b) phase $\angle M_V$. The measurement is compared to a simulation of the same quantity.

---

[9] A similar deviation is observed when comparing measurements and simulations of the electrical input impedance of the piezoelectric disk (not shown here).

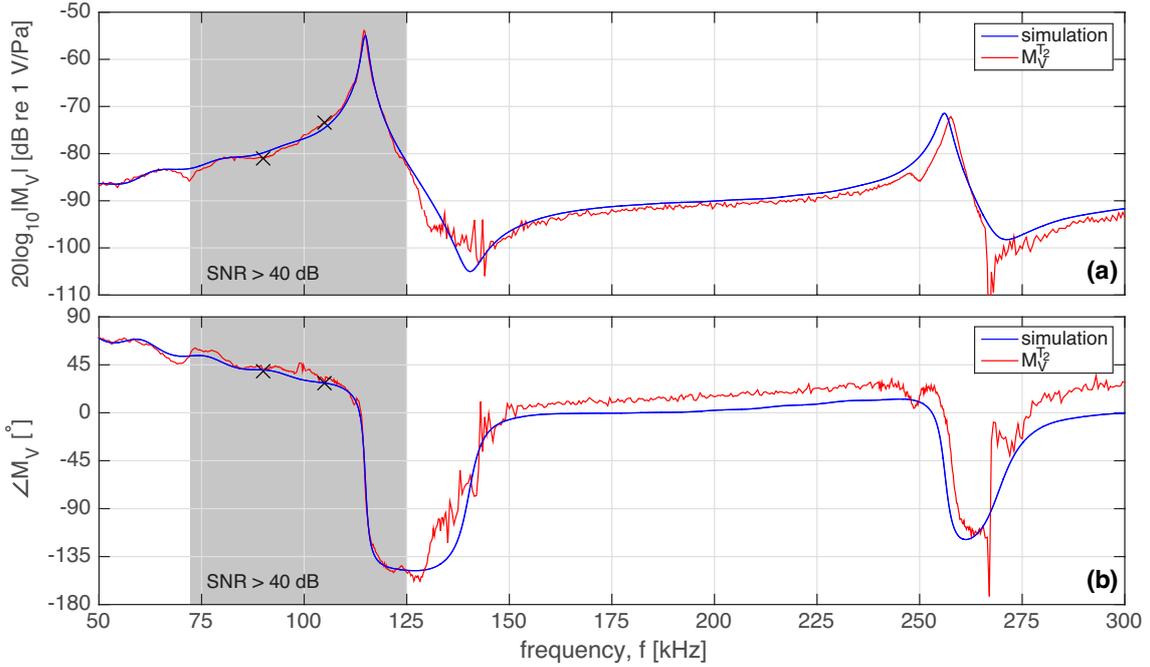

Figure 6: (a) Magnitude $|M_V|$ and (b) phase $\angle M_V$ of the receiving voltage sensitivity, $M_V^{T_2}$. The grey area indicates the frequency range where a SNR > 40 dB is achieved.

In Fig. 6 (a) around 70–125 kHz a fair correspondence between the measurement and the simulation is observed; the observed deviations are less than 2 dB. The "dip" in the measurement at 250 kHz is due to non-linearity, and the deviation between the measurement and simulation at the peak around 255–260 kHz is partly explained by the material constants used in the FE-simulations and by non-linear effects (small). At 200 kHz the deviation between the measurement and simulation is less then 1 dB; at 300 kHz the deviation is less then 2 dB.

In Fig. 6 (b) around 70–125 kHz a fair correspondence between the measurement and the simulation is observed; the observed deviations are less then 10°. The "dip" in the measurement at 250 kHz is due to non-linearity. As in Fig. 4 (b) and Fig. 5 (b) the deviation between the measurement and simulation increase with increasing frequency. At 200 kHz the deviation is approximately 15°; at 300 kHz the deviation is approximately 25°. The deviation is hypothesized to stem from the same sources listed in Sec. 5.1.

For both Fig. 6 (a) and (b) the large deviation around 260–275 kHz is attributed to diffraction effects [29].

### 5.4 Measurement uncertainty

In Fig. 7 the estimated relative measurement uncertainties [29]

$$u_r(|M_V|) = \frac{u(|M_V|)}{|M_V|},$$
$$u_r(|H_{15open}^{VV}|) = \frac{u(|H_{15open}^{VV}|)}{|H_{15open}^{VV}|}, \quad (10)$$

are given. The red crosses indicates the interval where the 1 V excitation voltage has been used. Since $u_r(|S_V|) = u_r(|M_V|)$ this is not shown. The uncertainties are shown given a

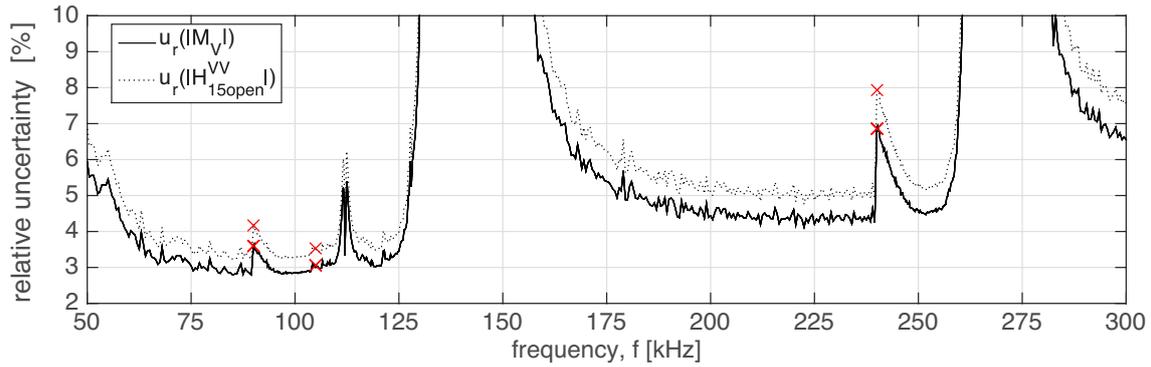

Figure 7: Relative measurement uncertainty given in percent for the magnitude of $M_V$ and $H^{VV}_{15open}$.

68.3% confidence level.

Noteworthy is that the uncertainty tends to about 3% in the range 70–125 kHz. This corresponds to the measurement uncertainty of the measurement amplifier and filter (not shown here). The peak at about 112–115 kHz is due to possible lack of steady-state conditions and electromagnetic cross-talk [29] [10]. Above 125 kHz the large uncertainty is due to poor SNR due to low sound pressure. It is also worth noticing that $u_r(|M_V|)$ is below 5% in the range 180–240 kHz, and 245–255 kHz. This indicates that it might be possible to perform calibration with adequate SNR also in this frequency range.

A Type A uncertainty was calculated for $\theta_6$, and this was found to be less than 3% in the range 70–125 kHz given a 68.3% confidence level. The Type A uncertainty was obtained by 6 repeated measurements [29].

# 6 Conclusion

A modified conventional three-transducer reciprocity calibration method is used to characterize 20 x 2 mm circular piezoelectric ceramic disks radiating in air at 1 atm. Preliminary results are shown for the magnitude and phase responses of the transmitting voltage response and the free-field open-circuit receiving voltage sensitivity of such transducers. Frequencies up to 300 kHz are investigated, covering the first and second radial modes of the disks.

Finite element simulations of the measurement system and the measured frequency responses are used to support the measurements, interpret the results, and remove a 360° ambiguity in the measured phase response at low frequencies, at which poor signal-to-noise ratio (SNR) is experienced. A fair agreement between measured and simulated magnitude and phase responses is obtained over the frequency band up to 300 kHz.

Promising characterization results are obtained for the magnitude and phase responses in a frequency band 75-125 kHz around the first radial mode at 100 kHz, in which SNR > 40 dB is achieved.

Factors limiting the measurements include low signal-to-noise ratio (SNR) due to coherent electromagnetic noise, acoustic reflections in the measurement system, non-linearity in the piezoelectric material, and absence of true far field conditions (near field diffraction effects). A laser sensor system has been implemented and calibrated to improve distance measurements and transducer alignment, and thus to improve measurement reproducibility and control. A more detailed analysis and additional results can be found in [29].

---

[10]Significant electromagnetic cross-talk is observed when the piezoelectric disks are operated with exposed electrodes.

Further work is needed to demonstrate and quantify the accuracy obtained using the three-transducer reciprocity calibration method, and to achieve transducer characterization at a calibration accuracy level. Further work may also address extension of the method to characterization above the calibrated frequency range of the B&K 4138 microphone. Construction of more sensitive piezoelectric transducers for gas, with reduced nonlinear effects and possibilities for improved shielding with respect to electromagnetic cross-talk, are expected to be important in these perspectives.

## 7 Acknowledgments

The current work is performed as part of a master project by the first author [29], supported by the Michelsen Centre for Industrial Measurement Science and Technology, Bergen, Norway.

# Appendix A  Derivation of the reciprocity calibration equations

In the current section the derivation of $M_V^{T_2}$ and $S_V^{T_1}$ will be presented. These equations will be referred to as the reciprocity calibration equations. The derivations are based upon [34], which are similar to the equations used in e.g. [7]. The derivation in this section differ from both due to the wave number $k = 2\pi f/c$, where $c$ is the speed of sound. The speed of sound is dependent on several environmental parameters: temperature, relative humidity, ambient pressure and $CO_2$ concentration [44]. The calibration by the reciprocity method utilizes three measurements where each measurement is associated with a $k$. In general, the three $k$'s will differ in value. Incorrect handling of the three $k$'s might therefore lead to wrong phase values.

## A.1  Receiving voltage sensitivity, $M_V$

From Fig. 1 three voltage-to-voltage transfer functions, $H_{15open}^{VV}$, are identified:

$$H_{15open}^{VV(1)} \equiv \frac{V_{5open}^{(1)}}{V_1^{(1)}} = M_V^{T_2(1)} S_V^{T_1(1)} \frac{d_0}{d_1} e^{ik^{(1)}(d_0-d_1)}, \tag{11}$$

$$H_{15open}^{VV(2)} \equiv \frac{V_{5open}^{(2)}}{V_1^{(2)}} = M_V^{T_3(2)} S_V^{T_1(2)} \frac{d_0}{d_2} e^{ik^{(2)}(d_0-d_2)}, \tag{12}$$

$$H_{15open}^{VV(3)} \equiv \frac{V_{5open}^{(3)}}{V_1^{(3)}} = M_V^{T_2(3)} S_V^{T_3(3)} \frac{d_0}{d_3} e^{ik^{(2)}(d_0-d_3)} \tag{13}$$

where the superscripts refer to what measurements the transfer functions are obtained from. Dividing Eq. (11) on (12) and solving for $M_V^{T_2(1)}$ yields

$$M_V^{T_2(1)} = M_V^{T_3(2)} \frac{H_{15open}^{VV(1)}}{H_{15open}^{VV(2)}} \frac{d_1}{d_2} e^{-ik^{(1)}(d_0-d_1)} e^{ik^{(2)}(d_0-d_2)}. \tag{14}$$

The spherical reciprocity parameter, $J$, is defined as the ratio of receiving voltage sensitivity to the transmitting current response [55], i.e.:

$$J \equiv \frac{M_V}{S_I} = \frac{M_V}{S_V Z_T}, \tag{15}$$

where $Z_T$ is the impedance of the transmitting transducer. Solving Eq. (15) for $S_V$ while applying correct notation, yields

$$S_V^{T_3(3)} = \frac{M_V^{T_3(3)}}{J^{(3)} Z_{T_3}}, \tag{16}$$

where the quantities adopt the environmental dependencies from measurement three since $S_V^{T_3(3)}$ in Eq. (16) will replace $S_V^{T_3(3)}$ in Eq. (13), i.e.:

$$H_{15open}^{VV(3)} = M_V^{T_2(3)} \frac{M_V^{T_3(3)}}{J^{(3)} Z_{T_3}} \frac{d_0}{d_3} e^{ik(d_0-d_3)}. \tag{17}$$

Solving Eq. (17) for $M_V^{T_3(3)}$, inserting the result in Eq. (14) and solving for $M_V^{T_2}$, yields

$$M_V^{T_2} = \left[ J^{(3)} Z_{T_3} \frac{H_{15open}^{VV(1)} H_{15open}^{VV(3)}}{H_{15open}^{VV(2)}} \frac{d_1}{d_0} \frac{d_3}{d_2} e^{-ik^{(1)}(d_0-d_1)} e^{ik^{(2)}(d_0-d_2)} e^{-ik^{(3)}(d_0-d_3)} \right]^{\frac{1}{2}}. \quad (18)$$

### A.2 Transmitting voltage response, $S_V$

Dividing Eq. (11) on (13) and solving for $S_V^{T_1(1)}$ yields

$$S_V^{T_1(1)} = S_V^{T_3(3)} \frac{H_{15open}^{VV(1)}}{H_{15open}^{VV(3)}} \frac{d_1}{d_3} e^{-ik^{(1)}(d_0-d_1)} e^{ik^{(3)}(d_0-d_3)}. \quad (19)$$

Use of Eq. (16) to replace $S_V^{T_3(3)}$ in Eq. (19), yields

$$S_V^{T_1(1)} = \frac{M_V^{T_3(3)}}{J^{(3)} Z_{T_3}} \frac{H_{15open}^{VV(1)}}{H_{15open}^{VV(3)}} \frac{d_1}{d_3} e^{-1k^{(1)}(d_0-d_1)} e^{ik^{(3)}(d_0-d_3)}. \quad (20)$$

Solving Eq. (12) with respect to $M_V^{T_3(2)}$, inserting the result in Eq. (20) and solving for $S_V^{T_1}$ yields

$$S_V^{T_1} = \left[ \frac{1}{J^{(3)} Z_{T_3}} \frac{H_{15open}^{VV(1)} H_{15open}^{VV(2)}}{H_{15open}^{VV(3)}} \frac{d_1}{d_0} \frac{d_2}{d_3} e^{-ik^{(1)}(d_0-d_1)} e^{-ik^{(2)}(d_0-d_2)} e^{ik^{(3)}(d_0-d_3)} \right]^{\frac{1}{2}}. \quad (21)$$

## Appendix B  Measurement set-up

### B.1 Electromechanical quantities of the measurement set-up

Table 1: Electromechanical quantities of the measurement set-up.

| Quantity | Description |
|---|---|
| $V_0 = V_0(f)$ | output voltage at the terminals of the function generator |
| $V_{0m} = V_{0m}(f)$ | recorded voltage given the signal transmission through cable 2 |
| $V_1 = V_1(f)$ | input voltage at the terminals of the transmitting disk |
| $u_2(r = 0, f)$ | particle displacement at the center of the face of the transmitting disk |
| $p_3 = p_3(d_0, f)$ | on-axis free- and far-field sound pressure at a ref. distance $d_0 = 1$ m |
| $p_4 = p_4(r, d, f)$ | free-field sound pressure at a separation distance $z = d$ |
| $V_5 = V_5(f)$ | output voltage at the terminals of the receiving disk |
| $V_{5'} = V_{5'}(f)$ | input voltage at the terminals of the amplifier |
| $V_{6'} = V_{6'}(f)$ | output voltage at the terminals of the amplifier |
| $V_6 = V_6(f)$ | input voltage at the terminals of the oscilloscope |
| Quantities not readily visible in Fig. 2 | |
| $V_{5open} = V_{5open}(f)$ | open-circuit output voltage at the terminals of the receiving disk |
| $V_{6open} = V_{6open}(f)$ | open-circuit output voltage at the terminals of the amplifier |
| $V_{gen} = V_{gen}(f)$ | open-circuit generator voltage |

In Table 1 $(r, z)$ are the coordinates in a cylindrical coordinate system [11], where $r = \sqrt{x^2 + y^2}$, and $(x, y, z)$ are the Cartesian coordinate axes.

## B.2 Corrections

Table 2: Corrections used in Eq. (4).

| Quantity | Correction accounting for: | Estimation method |
| --- | --- | --- |
| $H_{0m1}^{VV} \equiv \frac{V_1}{V_{0m}}$ | Oscilloscope input impedance, cable 1, cable 2, and transmitter electrical impedance | Transmission line model calculation, measurement of transmitter input electrical impedance |
| $H_{5open5'}^{VV} \equiv \frac{V_{5'}}{V_{5open}}$ | Receiver electrical impedance, cable 3 and amplifier input impedance | Transmission line model calculation, measurement of receiver input electrical impedance |
| $H_{5'6open}^{VV} \equiv \frac{V_{6open}}{V_{5'}}$ | Amplifier and filter | Measurement |
| $H_{6open6}^{VV} \equiv \frac{V_6}{V_{6open}}$ | Amplifier output impedance, cable 4 and oscilloscope input impedance | Transmission line model calculation |
| $C_\alpha$ | Attenuation | Calculation by ANSI S1.26 [56] |
| $C_{dif}$ | Diffraction | FE-simulation |

# Appendix C   Phase of a time-shifted Fourier transform signal

In Fig. 8 a schematic of a sinusoidal signal, $V_6(t)$ [12], that is time-shifted with a constant $t_0$ is given. In the same figure, a FFT-window is indicated, where the lower bounds of the FFT-window is placed in $t_p$. A discrepancy between the actual onset of the time-shifted signal and the position of the FFT-window's lower bound is seen. This time-discrepancy is denoted $\Delta t = t_0 - t_p$. Two examples will be presented: 1) $\Delta t = 0$, and 2) $\Delta t > 0$. The first example illustrates the phase contributions given that the FFT-window's lower bound is placed in the actual signal onset, whereas the second example illustrates why the FFT-window's lower bound has to be placed within one signal period of the actual signal onset.

The case where $\Delta t = 0$ will be considered first. In [43] it is shown that the Fourier transform of a time-shifted signal, $V_6(t - t_p)$, can be solved substituting $s = t - t_p$, i.e.

---

[11]Note that due to the axisymmetry of the system the azimuthal ($\theta$) is suppressed from the cylindrical coordinates, $(r, \theta, z)$.

[12]The notation $V_6$ will be used in the examples in the current section, however the examples that will be given are of a time-shifted sinusoid, i.e. no phase distortions in e.g. a transmitter or receiver exist.

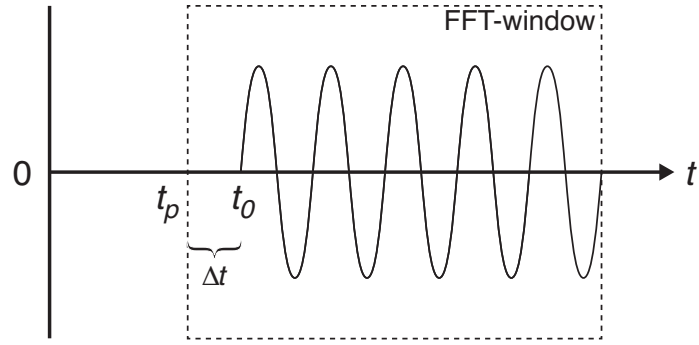

Figure 8: Schematic of a sinusoidal signal time-shifted with a constant $t_0$. The FFT-window's lover bound is placed in $t_p$.

$$\begin{aligned}
\int_{-\infty}^{\infty} V_6(t-t_p)e^{-i2\pi ft}dt &= \int_{-\infty}^{\infty} V_6(s)e^{-i2\pi f(s+t_p)}ds \\
&= e^{-i2\pi ft_p} \int_{-\infty}^{\infty} V_6(s)e^{-i2\pi fs}ds \\
&= e^{-i2\pi ft_p} V_6(f) \\
&= |V_6(f)|e^{i(\theta_6^{slow}-2\pi ft_p)}
\end{aligned} \quad (22)$$

where $2\pi ft_p$ is a linear phase dependent on frequency, $f$, and the time-shift constant $t_p$, and $\theta_6^{slow}$ is the phase of the sine wave. The phases can be expressed without the exponential notation as

$$\begin{aligned}
\theta_6 &= \theta_6^{slow} - 2\pi ft_p \\
&= \theta_6^{slow} - 2\pi ft_0,
\end{aligned} \quad (23)$$

since $t_p = t_0$. In the current example only a time-shifted sine wave is considered. However, if a signal is propagating through physical measurement equipment phase distortions will generally occur in the equipment. These phase distortions will then be associated with $\theta_6^{slow}$ since the position of the FFT-window remains the same.

The case where $\Delta t > 0$ will now be considered. The time-shifted signal can now be expressed as $V(t - t_p - \Delta t)$, and $s = t - t_p - \Delta t$. The derivation will be omitted, but the phases can be expressed as:

$$\begin{aligned}
\theta_6 &= \theta_6^{slow} - 2\pi f(t_p + \Delta t) \\
&= \theta_6^{slow} - 2\pi ft_0,
\end{aligned} \quad (24)$$

which is the same results as in Eq. (23). That is, the phase $\theta_6$ is independent of the placement of the FFT-window as long as $\Delta t < T$, where $T = 1/f$ is signal period. If $\Delta t > T$ phase offsets equal to $N \cdot 2\pi$, where $N$ is an integer, will occur. To understand this one needs to remember that the phase values of a Fourier transform are confined to $\pm \pi$, and recollect that both phases $\theta_6^{slow}$ and $2\pi f\Delta t$ are obtained by the Fourier transform. Formally this can be stated as

$$-\pi \leq \angle e^{i(\theta_6^{slow}-2\pi f\Delta t)} \leq \pi, \quad (25)$$

where $\angle$ denotes the phase angle.

## Appendix D  Equations for the speed of sound

The sound speed model proposed by Cramer takes into account the temperature in Kelvin, $T_K$, atmospheric pressure, $p$, humidity and $CO_2$ concentration, and is given in [44] as

$$c_0^2 = \gamma \frac{RT_K}{M}\left(1 + \frac{2pB}{RT_K}\right), \qquad (26)$$

where $c_0$ is the zero frequency speed of sound, $\gamma = C_P/C_V$ is the specific heat ratio where $C_P$ and $C_V$ are the specific heat at constant pressure and volume, respectively, $R$ is the universal gas constant, $M$ is the molecular mass and $B$ is the second virial coefficient of state.

The expression in Eq. (26) can be expanded to account for dispersion due to the vibrational relaxation effects of oxygen and nitrogen, both of which are functions of frequency and are regarded as the greatest contributors to the absorption of sound by the atmosphere [57]. The speed of sound can then be estimated using a model, proposed by Morfey and Howell [45], which takes into account dispersion:

$$\frac{1}{c_0} - \frac{1}{c} = \frac{\alpha_{vN}}{2\pi f_{rN}} + \frac{\alpha_{vO}}{2\pi f_{rO}}, \qquad (27)$$

where $c = c(f)$ is the estimated speed of sound at a frequency $f$ including effects of dispersion, $\alpha_{vN}$ and $\alpha_{vO}$ are the plane wave attenuation coefficients due to vibrational relaxation of nitrogen and oxygen, respectively, and $f_{rN}$ and $f_{rO}$ are the relaxation frequencies for nitrogen and oxygen, respectively.